\input phyzzx
\baselineskip 24pt plus 1pt minus 1pt
\hfill\vbox{\hbox{UUPHY/96/02}
\hbox {August, 1996}
\hbox {hep-th/9608028}}\break
\NPrefs
\let\refmark=\NPrefmark
\def\define#1#2\par{\def#1{\Ref#1{#2}\edef#1{\noexpand\refmark{#1}}}}
\def\con#1#2\noc{\let\?=\Ref\let\<=\refmark\let\Ref=\REFS
         \let\refmark=\undefined#1\let\Ref=\REFSCON#2
         \let\Ref=\?\let\refmark=\<\refsend}

\define\DV
H. J. de Vega and N. Sanchez, Phys. Lett. {\bf B197} (1987) 320;
Nucl. Phys. {\bf B309} (1988) 552, 577.

\define\COSMIC
H. J. de Vega and N. Sanchez, Phys. Rev. {\bf D42} (1990) 3969;
H. J. de Vega, M. Ramon-Mendrano and N. Sanchez, Nucl. Phys. 
{\bf B374} (1992) 405.

\define\DESITTER
H. J. de Vega and N. Sanchez, Phys. Rev. {\bf D47} (1993) 3394.

\define\GRWAVE
H. J. de Vega and N. Sanchez, Phys. Lett. {\bf B244} (1990) 215;
Phys. Rev. Lett. {\bf 65} (1990) 1517; Int. J. Mod. Phys. {\bf A7}
(1992) 3043; Nucl. Phys. {\bf B317} (1989) 706; {\bf B317} (1989) 731;
D. Amati and C. Klimcik, Phys. Lett. {\bf B210} (1988) 92.

\define\EQUIFRO
V. P. Frolov, V. D. Skarzhinsky, A. I. Zelnikov and O. Heinrich, 
Phys. Lett. {\bf B224} (1989) 255.

\define\FROLOV
C. O. Lousto and N. Sanchez, Phys. Rev. {\bf D47} (1993) 4498;
V. P. Frolov and A. L. Larsen, Nucl. Phys. {\bf B449} (1995) 149.

\define\COSMOL
N. Sanchez and G. Veneziano, Nucl. Phys. {\bf B333} (19900 253.

\define\SAYAN
S. Kar, Phys. Rev. {\bf D52} (1995) 2036.

\define\MULTI
H. J. de Vega, A. V. Mikhailov and N. Sanchez, Teor. Math. Fiz. 
{\bf 94} (1993) 232; F. Combes, H. J. de Vega, A. V. Mikhailov
and N. Sanchez, Phys. Rev. {\bf D50} (1994) 2754; H. J. de Vega, 
A. L. Larsen and N. Sanchez, Nucl. Phys. {\bf B427} (1994) 643.

\define\CIRCULAR
A. L. Larsen and N. Sanchez, Phys. Rev. {\bf D50} (1994) 7493.

\define\STATIONARY
A. L. Larsen and N. Sanchez, Phys. Rev. {\bf D51} (1995) 6929.

\define\FRO
A. L. Larsen and V. P. Frolov, Nucl. Phys. {\bf B414} (1994) 129.

\define\GUVEN
B. Carter, Phys. Rev. {\bf D48} (1993) 4835; J. Guven, Phys. 
Rev. {\bf D48} (1993) 5562; R. Capovilla and J. Guven, Phys. Rev.
{\bf D51} (1995) 6752.

\define\LARSEN
A. L. Larsen, Nucl. Phys. {\bf B412} (1994) 372.

\define\DYON
D. A. Lowe and A. Strominger, Phys. Rev. Lett. {\bf 73} (1994) 1468.

\define\POLCHINSKI
S. Giddings, J. Polchinski and A. Strominger, Phys. Rev. {\bf D48}
(1993) 5784.

\define\ORTIZ
A. Achhcarro and M. Ortiz, Phys. Rev. {\bf D48} (1993) 3600.

\define\BTZ
M. Banados, C. Teitelboim and J. Zanelli, Phys. Rev. Lett. {\bf 69}
(1992) 1849; G. T. Horowitz and D. L. Welch, Phys. Rev. Lett. {\bf 71}
(1993) 328.
 
\define\ABRAM
M. Abramowitz And I. Stegun, Handbook of Mathematical Functions 
(Dover, New York, 1965).

\define\GARRIGA
J. Garriga and A. Vilenkin, Phys. Rev. {\bf D44} (1991) 1007;
A. Vilenkin, Phys. Rep. {\bf 121} (1985) 264.


\title{\bf{STRING PROPAGATION IN FOUR-DIMENSIONAL DYONIC 
BLACK HOLE BACKGROUND}}
\author{Swapna Mahapatra\foot{e-mail: swapna@iopb.ernet.in}} 
\address{Department of Physics, Utkal University, Bhubaneswar-751004,
India.}

\abstract
We study string propagation in an exact, stringy, four-dimensional
dyonic black hole background. The general solutions describing string
configurations are obtained by solving the string equations of motion
and constraints. By using the covariant formalism, we also investigate
the propagation of physical perturbations along the string in the given 
curved background. 
\endpage
\noindent
{\bf I. Introduction}

The study of string dynamics in curved space-time started in \DV\ has 
become an interesting area of research in recent times. The classical 
string equations 
of motion and the constraints in curved space-time become 
highly nonlinear, coupled partial differential equations and hence
it is often difficult to find exact and complete solutions of these
equations. However, exact solutions to these equations of motion 
and constraints have been obtained in some specific curved 
space-times, such as, cosmic string backgrounds\COSMIC , black hole 
spacetimes\EQUIFRO\FROLOV, de-Sitter space-times\DESITTER, gravitational 
wave backgrounds\GRWAVE, cosmological background\COSMOL\ and wormhole 
background\SAYAN.
The string equations of motion and constraints are exactly integrable
in $D$-dimensional de-Sitter space-times\DESITTER. The novel feature of
strings in de-Sitter space-time is the appearance of multi-string solutions 
\MULTI, 
where there is only one single world-sheet, but infinitely many
different and independent strings. Also, by using certain 
appropriate ans${\ddot a}$tze, like circular string ansatz\CIRCULAR\ or 
stationary
string ansatz\STATIONARY, one can obtain a general family of exact solutions.
In stationary, axially symmetric space-times, the circular string 
ansatz corresponds to decoupling of the dependence of $x^{\mu}$
on the world   
sheet parameter $\sigma$ and consequently reduces the equations
of motion and the constraints to simpler coupled ordinary differential
equations. On the other hand, the stationary string ansatz
decouples the dependence of $x^{\mu}$ on the worldsheet coordinate 
$\tau$
and again the string equations of motion and constraints reduce
to separated first order ordinary differential equations, which 
are easier to handle with. It has also been noticed that consideration
of stationary strings in static or time-independent backgrounds
simplifies the problem a lot, where the stationary string 
configuration in static backgrounds can be described by geodesic 
equations in a certain three dimensional "unphysical" space\EQUIFRO. 
After 
knowing the extremal string configuration, one can also consider
small perturbations around the stationary strings. Larsen and 
Frolov\FRO\ have developed a covariant formalism for small 
perturbations
propagating along a string in curved space-time (see also\GUVEN)
and have subsequently analyzed stationary strings in quasi 
newtonian, Rindler, black holes and de-Sitter space-times. The 
second variation of Polyakov action essentially gives the equations
describing the propagation of perturbations along a stationary 
string in a static background. By doing the analysis in the properly
chosen three dimensional "unphysical" space-time, one obtains simple 
wave equations such as the P${\ddot o}$schl-Teller equation
in quantum mechanics for strings in Rindler and in de-Sitter
space-time. For Schwarzschild blackhole background, the
wave equation has been explicitly analysed by using the weak-field
expansion and a scattering formalism has also been set up\LARSEN.

In this paper, we analyze the equilibrium string configuration 
in the background of exact, four-dimensional dyonic blackhole
of Lowe and Strominger\DYON, 
which has been obtained by tensoring the two dimensional
electrically charged blackholes (these are related to $2 + 1$ 
dimensional rotating blackhole of Banados, Teitelboim and Zanelli    
\BTZ\ by Kaluza-Klein reduction) with two dimensional 
${SU(2)\over{Z(m)}}$ coset models. We make use of the stationary 
string ansatz and the solutions of string equations of motion are
obtained in terms of elliptic functions. However, we do not find 
any multistring solution in the above case. We also investigate 
the nature of perturbations as well as their propagation around
the string configuration.  

\noindent
{\bf II. Stationary string ansatz and four-dimensional 
dyonic blackhole background}

The family of exact, four-dimensional dyonic black holes in string 
theory \DYON\ are constructed
as a tensor product of electrically charged two-dimensional 
blackholes with the angular magnetic monopole CFT obtained 
by quotienting a $SU(2)$ WZW model by the discrete subgroup 
$Z(m)$ \POLCHINSKI, where $m$ is an integer. The level of the 
corresponding 
WZW model is denoted as $k_{SU}$. The two dimensional, electrically
charged black hole part is obtained by a Kaluza-Klein reduction
\ORTIZ of the string analogue of $2 + 1$ dimensional, rotating 
blackhole
solution\BTZ. This tensor product leads to a solution describing
the throat limit of a four-dimensional blackhole  with electric
and magnetic charge. The corresponding metric is given by \DYON,
$$
\eqalign{d s^2 &= -\left(-M + {r^2\over{l^2}} + 
{J^2\over{4 r^2}}\right)dt^2 
+ \left(-M + {r^2\over{l^2}} + 
{J^2\over{4 r^2}}\right)^{- 1}dr^2 + \cr   
& ~~~~~~{1\over 4} k_{SU} d\theta^2 + {1\over 4} k_{SU} sin^2\theta
d\phi^2\cr}\eqn\one$$
where $k_{SU}$ is the level of $SU(2)$ WZW model as discussed before,
$M$ is the mass of the blackhole, $J$ is the angular momentum and
the cosmological constant is proportional to $l^2$.

Next, we analyze the string configuration in the background
of above four-dimensional dyonic blackhole by using the stationary 
string ansatz. We consider strings in the equatorial plane for 
which $\theta = {\pi\over 2}$. The stationary string ansatz 
is given by,
$$
t = \tau, \qquad r = r(\sigma), \qquad \phi = \phi(\sigma)
\eqn\two$$
The string equations of motion and the constraints are given 
by,
$$
{\ddot x}^{\mu} - {x''}^{\mu} + \Gamma^{\mu}_{\rho\sigma}
(\dot x^{\rho}\dot x^{\sigma} - {x'}^{\rho}{x'}^{\sigma}) = 0
\eqn\thr$$
$$
g_{\mu\nu}\dot x^{\mu}{x'}^{\nu} = g_{\mu\nu}(\dot x^{\mu}
\dot x^{\nu} + {x'}^{\mu}{x'}^{\nu}) = 0\eqn\four$$
where dot and prime denote derivatives with respect to $\tau$
and $\sigma$ respectively and $\mu$, $\nu$ run over $t, r, \theta$ and
$\phi$. Since we consider strings in equatorial plane, we only 
have to determine the functions $r(\sigma)$ and $\phi(\sigma)$
by solving the string equations of motion and the constraints.
The metric of a static space-time can in general be written as,
$$ 
g_{\mu\nu} = \pmatrix{- F & 0\cr
0 & {H_{i j}\over F}\cr}\eqn\five$$
where $\partial_t F = 0;~~ \partial_t H_{i j} = 0$ and $i, j =
1, 2, 3$. $H_{i j}$ is the metric of the three-dimensional 
unphysical space. For the stationary string ansatz, the equations 
of motion for the Nambu-Goto action reduce to,
$$
{x''}^i +{\tilde\Gamma}^i_{j k} {x'}^j {x'}^k = 0\eqn\six$$
where ${\tilde\Gamma}^i_{j k}$ is the Christoffel symbol for the
metric $H_{i j}$. Equation \six  is nothing but the geodesic
equation in the three-dimensional unphysical space, with the 
line element,
$$
d {\tilde s}^2 = H_{i j} d x^i d x^j\eqn\sevn$$
and,
$$H_{i j} x'^i x'^j = 1\eqn\eit$$
For the four dimensional dyonic black hole background,
$$\eqalign{F &= -M + {r^2\over{l^2}} + {J^2\over{4 r^2}}\cr
H_{i j} &= diag (1,~~ \Delta, ~~\Delta\sin^2\theta)\cr}\eqn\nine$$
where,
$$\Delta = {1\over 4} k_{S U} \left(-M + {r^2\over{l^2}} + 
{J^2\over{4 r^2}}\right)\eqn\ten$$
The non-zero components of the Christoffel symbol 
${\tilde\Gamma}^i_{j k}$ for the metric in the "unphysical" space
are given by,
$$
\eqalign{{\tilde\Gamma^r_{\phi\phi}} & = 
-{k_{SU}\over 4}\left({r\over l^2} -
{J^2\over{4 r^3}}\right)\sin^2\theta \cr
& = {\tilde\Gamma^r_{\theta\theta}}
\sin^2\theta \cr
{\tilde\Gamma}^{\theta}_{r\theta} &= {\tilde\Gamma}^{\phi}_{r\phi}\cr
&= {1\over F}\left({r\over l^2} - {J^2\over{4 r^3}}\right)\cr
{\tilde\Gamma}^{\theta}_{\phi\phi} & = - \cos\theta\sin\theta\cr
{\tilde\Gamma}^{\phi}_{\phi\theta} & = \cot\theta\cr}
\eqn\eleven$$
The non-vanishing components of Riemann tensor are,
$$
\eqalign{{\tilde R}_{r\phi r\phi} &= 
\sin^2\theta{\tilde R}_{r\theta r\theta}\cr
&= \sin^2\theta {k_{SU}\over 4 F} \left({M\over l^2} +
{{3 J^2 M}\over{4 r^4}} - {{3 J^2\over{2 r^2 l^2}}} - {J^4\over
{8 r^6}}\right)\cr
{\tilde R}_{\theta\phi\theta\phi} &= {k_{S U}\over 4}\sin^2\theta
\left[F - {k_{S U}\over 4} {\left({r\over l^2} - 
{J^2\over{4 r^3}}\right)}^2 \right]\cr} \eqn\twelv$$

For strings in the equatorial plane ($\theta = {\pi\over 2}$), the
geodesic equations are given by,
$$
{\phi''} + 2 {\tilde\Gamma}^{\phi}_{r\phi} r' \phi' = 0\eqn\thirtn$$
$$
r'' + {\tilde\Gamma}^r_{\phi\phi} {\phi'}^2 = 0\eqn\fourtn$$

The constraint equation \eit ~is given by,
$$
{r'}^2(\sigma) + \Delta{\phi'}^2(\sigma) = 1\eqn\fiftn$$
The string configuration is known by solving equations \thirtn,
\fourtn ~and \fiftn. Integrating the $\phi$ equation \thirtn,
one obtains, 
$$\phi'({\sigma}) = {b\over\Delta}\eqn\sixtn$$
Integrating equation \fourtn ~using \sixtn, one obtains,
$$
r'(\sigma) = \pm{\sqrt{1 - {b^2\over\Delta}}}\eqn\seventn$$
where $b$ is an integration constant. These are the two first 
order differential equations, whose solutions will give the
string configuration. The $r$ equation of motion can
also be written as,
$$
r'^2 + V(r) = 0.\eqn\eitn$$
where, $V(r)$ is the effective potential given by,
$$
V(r) = - (1 - {b^2\over\Delta})\eqn\nintn$$
Qualitatively the possible string configurations can also be known 
by studying the zeros of 
the potential. The solution of equations \sixtn ~and \seventn ~ 
are obtained in terms of elliptic functions. The world sheet 
coordinate $\sigma$ can be replaced in terms of conformal string 
parameter $\sigma_c$, so that $d\sigma_c = {d\sigma\over F}$. 
Hence we obtain,
$$
r'(\sigma_c) = \pm{4\Delta\over{k_{S U}}}{\sqrt{1 - {b^2\over\Delta}}}
\eqn\twenty$$
$$
\phi'(\sigma_c) = {4 b\over{k_{S U}}}\eqn\twentyone$$
where, ${4\Delta\over{k_{S U}}} = - M + {r^2\over{l^2}}
+ {J^2\over{4 r^2}}$. These two equations are difficult to solve
for nonzero $J$. Hence, we restrict ourselves to the case when 
$J = 0$ and also take $M = 1$. In this case, the solution for 
$r^2(\sigma_c)$ is obtained in terms of elliptic functions,
$$
r^2(\sigma_c) = l^4\wp (\sigma_c + z_0; g_2, g_3) + {2 l^2\over 3}
\left(1 + {2 b^2\over{k_{S U}}}\right)\eqn\twentytwo$$
where, $\wp(z)$ is the Weierstrass elliptic $\wp$-function \ABRAM\ with 
the variants $g_2$ and $g_3$ taking the value,
$$
\eqalign{g_2 &= {4\over{l^6}}\left({l^2\over 3} + 
{4 b^2 l^2\over{3 k_{S U}}} + {16\over 3} 
{b^4 l^2\over{k_{S U}^2}}\right)\cr
g_3 &= -{4\over{3 l^8}}\left({2 l^2\over 9} + {4\over 3}{b^2 l^2
\over{k_{S U}}} - {16\over 3}{b^4 l^2\over{k_{S U}^2}} - 
{128\over 9}{b^6 l^2\over{k_{S U}^3}}\right)\cr}\eqn\twentythr$$
Here $z_0$ is an integration constant. The discriminant $\Delta$ is 
given by,
$$
\Delta = {16\beta^2\over{l^{12}}} (1 + \beta)^2
\eqn\twentyfour$$
where, $\beta = {4 b^2\over{k_{S U}}}$. Since $\Delta > 0$,
all the roots $e_1$, $e_2$, $e_3$ are real. The roots are given by,
$$
\eqalign{e_1 & = {1\over{3 l^2}} (1 + 2\beta)\cr
e_2 &= {1\over{3 l^2}} (1 - \beta) \cr
e_3 &= -{2\over 3 l^2}(1 + {\beta\over 2})\cr}\eqn\twentyfive$$
where, $e_1 > e_2 > e_3$. We can also express the solution
\twentytwo ~in terms of Jacobi's elliptic function, which
is given by,
$$
r^2(\sigma_c) = l^2\mu^2 ns^2[{\mu\over l}(\sigma_c + z_0), k]
\eqn\twentysix$$
where, $\mu^2 = 1 + \beta$ and $k$ is the elliptic modulus. This 
solution can be compared with strings in the background of $2 + 1$
dimensional blackhole anti-de-Sitter space-time (for $J = 0$ and
$M = 1$) given by \CIRCULAR,
$$
r^2(\sigma) = l^4\wp(\sigma - \sigma_0; g_2, g_3) + {2\over 3} l^2
\eqn\twentyseven$$
with variants,

$$\eqalign{g_2 &= {4\over{l^6}}\left({l^2\over 3} + L^2\right) \cr
g_3 &= -{4\over{3 l^8}}\left({2 l^2\over 9} + 
L^2\right)\cr}\eqn\twentyeit$$
where $L$ is an integration constant. The $\phi'(\sigma_c)$ 
equation \twentyone is simple and the solution is given by,
$$ 
\phi(\sigma_c) = \left({4 b\over{k_{S U}}}\right)\sigma_c + 
constant.\eqn\twentynine$$
which means that $\sigma_c$ is periodic with period 
${\pi k_{S U}\over{2 b}}$. In this case, there is no multistring 
solution possible, whereas for strings in the background of $2 + 1$ 
dimensional blackhole anti-de-Sitter space-time, one does obtain  
multistring solutions.  

\noindent
{\bf III. Straight string and circular string configuration}

Here we obtain a simple string configuration by choosing the 
parameter $b = 0$. In this limit, for the case of $J = 0$
$M = 1$, the two ordinary differential equations reduce to,
$$
\eqalign{r'^2(\sigma_c) &= ({r^2\over l^2} - 1)^2\cr
\phi'(\sigma_c) &= 0\cr}\eqn\thirty$$

Now the solution for the radial coordinate is obtained in terms of 
hyperbolic function
given by,
$$r^2(\sigma_c) = l^2 \tanh^2({\sigma_c\over l})\eqn\thirtyone$$
and,
$$\phi(\sigma_c) = constant\eqn\thirtytwo$$
This is nothing but a straight string configuration.  

The circular string configuration is given by $r = constant$,
$\phi$ as a periodic function of $\sigma$ and $\theta = {\pi\over 2}$.
Let us consider the case when ${4 b^2\over
k_{S U}} = 1$. In this case, the solutions of equations \twenty ~and
\twentyone ~reduce to 
$$
\eqalign{r^2(\sigma_c) &= l^2 , 2 l^2\cr
\phi(\sigma_c) &= {\sigma_c\over b}\cr}\eqn\thirtythree$$
Though they satisfy the two first order differential equations, 
the analysis of the zeros of the effective potential shows that 
they will not lead to circular string configuration. The effective 
potential is given by,
$$
V(r(\sigma_c)) = - ({r^2\over l^2} - 1) \left({r^2\over l^2} - 1
- {4 b^2\over{k_{S U}}}\right)\eqn\thirtyfour$$
This implies that the potential vanishes at 
$r = l$ and at $r = r_0 = l{[1 + {4 b^2\over{k_{S U}}}]}^{1\over 2}$.
For ${4 b^2\over{k_{S U}}} = 1$, $r$ takes the value,
$r = r_0 = {\sqrt 2} l$. But they do not lead to circular string
confuguration as they do not satisfy the condition ${{\partial V
\over{\partial r}}\mid}_{r = l, {\sqrt 2}l} = 0$, which is necessary 
for the original second order string equations of motion \thr ~and the 
constraints \four (or equivalently eqns. \thirtn, \fourtn ~and \fiftn) 
to be satisfied. On the other hand, a circular string configuration
is possible for non-zero $J$. Consider a circular string ansatz,

$$
t = \tau;\qquad r = C_1;\qquad \phi = {\sigma\over C_2}; ~~~
\theta = {\pi\over 2}\eqn\thirttyfive
$$
With this ansatz, eqn. \thirtn ~is trivially satisfied. Eqn. \fourtn
~determines $r$ as,
$$
r = C_1 = {\sqrt{J l\over 2}}\eqn\thirsix$$
Then eqn. \fiftn ~determines the constant $C_2$ as,
$$
C_2 = {1\over 2}{\sqrt{k_{S U} \left ({J\over l} - M\right )}}
\eqn\thirseven$$
Hence, $r = {\sqrt{J l\over 2}}$ and $\phi = {2\sigma\over{\sqrt{
k_{S U} \left ({J\over l} - M \right )}}}$, 
$\theta = {\pi\over 2}$, 
$t = \tau$ is an acceptable circular string solution provided 
$J > M l$. However, for the
horizons to exist, we need the condition $\Delta = 0$ and this 
implies that the solutions for $r^2$ are given by,
$$
r^2 = {1\over 2}\left [M l^2 \pm l {\sqrt{M^2 l^2 - J^2}}\right ]   
$$ 
As one can see from the above expression that there is no real root 
if $J > M l$. So this limit essentially gives us a geometry with 
no horizon. In fact, for stationary black hole solutions, which are 
not spherically symmetric, there can not be any circular string 
configuration. This can be seen in the following way.
Consider the general non spherically symmetric
metric of the form,
$$
d s^2 = - a(r) d t^2 + {d r^2\over{a(r)}} + B (d\theta^2 + \sin^2\theta
d\phi^2)\eqn\foorty$$
where $B$ is a constant. For the dyonic black hole metric, 
$a(r) = (- M + {r^2\over l^2} + {J^2\over{4 r^2}})$ and $B = {1\over 4}
k_{S U}$. Let us consider the circular string ansatz as discussed 
before. Now eqn. \thirtn ~is trivially satisfied. From eqn. \fourtn, 
we obtain the condition $a'(r) = 0$ (this will determine $r$ for 
specific solutions). Now, from the constraint equation \fiftn, ~  
one 
obtains, $C_2 = {\sqrt{B~ a(r)}}$. The existence of horizon implies
the condition $a(r) = 0$, which means that the constraint equation is
no longer satisfied. Hence, there can not be any circular string 
configuration if the geometry in the above case has a horizon. 
The possibility of a circular string 
is also ruled out in the case of Schwarzschild black hole \STATIONARY.
However, one obtains closed string configurations for wormhole 
geometry \SAYAN\ and de-Sitter space-time \FRO.  

\noindent
{\bf IV. Propagation of perturbation along the string}

The stability properties of the above obtained solutions can be 
investigated by studying the propagation of small perturbations along 
the strings in curved space-time. Propagation of small waves 
on a membrane and cosmic strings have been considered before\GARRIGA.
The equations of motion for small perturbations along the strings 
in arbitrary curved space-time can be obtained by the second 
variation of the Polyakov action. The first variation of the action 
gives the
equations of motion for the strings in curved space-time itself. 
The variations of equations of motion and constraints for strings
also gives the propagation of perturbations for the corresponding
string configurations. A covariant formalism for the analysis of 
perturbation has been developed by Larsen and Frolov \FRO\ and in
ref.\GUVEN. 

Let $x^{\mu}$ be a solution of equations \thr and \four.
Introducing two vectors $n^{\mu}_R$ ($R = 2, 3$) normal to 
the string world sheet, the perturbation can be decomposed as,
$$
\delta x^{\mu} = \delta x^R n^{\mu}_R + 
\delta x^A x^{\mu}_{, A}\eqn\thirtyfive$$
where, $A = 0, 1$ denote the string world sheet coordinates 
$\tau$ and $\sigma$ respectively and coma denotes partial 
differentiation. The normal vectors satisfy the conditions,
$$
g_{\mu\nu} n^{\mu}_R n^{\nu}_S = \delta_{R S}; \qquad
g_{\mu\nu} x^{\mu}_{, A} n^{\nu}_R = 0\eqn\thirtysix$$

Since the second term in the variation equation \thirtyfive ~ 
leaves the action invariant due to reparametrization invariance,
we will consider the first term in equation \thirtyfive, 
${\it i. e}$ $\delta x^{\mu} = \delta x^R n^{\mu}_R$. The effective
action for the physical perturbation is obtained in terms of
the second fundamental form and the normal fundamental form \FRO.
In the stationary string ansatz, where $t = x^0 = \tau$ and 
$x^i = x^i(\sigma)$, we have ,
$$
x^{\mu}_{, 0} = (1, 0, 0, 0) ; \qquad ~~~ x^{\mu}_{, 1} = 
(0, x'^i)\eqn\thirtyseven$$
Therefore, the components of the normal vectors can be written as,
$n^{\mu}_R = (0, n^i_R)$. 

$(x'^i, n_2^i, n_3^i)$ form an orthogonal system in the 
three dimensional "unphysical" space with the metric $H_{i j}$.
By defining, ${n^i_R\over{\sqrt F}} = {{\tilde n}_R}^i$, we
have the obvious relations,

$$\eqalign{H_{i j} {{\tilde n}_R}^i {{\tilde n}_S}^j &= {\delta}_{R S}\cr
H_{i j} x'^i {{\tilde n}_R}^j &= 0\cr}\eqn\thirtyeit$$

The equations of motion for the physical perturbations are 
given by,
$$
(\partial_{\sigma_c}^2 - \partial_{\sigma_c}^2)\delta x_R
= U_{R S} \delta x_S\eqn\thirtynine$$
where, $U_{R S}$ is the matrix potential given by,
$$
U_{R S} = V \delta_{R S} + F^{- 1} V_{R S}\eqn\forty$$
and,
$$
V  = {3\over{4 F^2}} \left({d F\over{d\sigma_c}}\right)^2 - 
{1\over{2 F}} {d^2 F\over{{d\sigma_c}^2}}\eqn\fortyo$$ 

$$
V_{R S} = {x'}^i(\sigma_c) {x'}^j(\sigma_c) {\tilde R}_{i j k l}
n_R^k n_s^l\eqn\fortyone$$
where, ${\tilde R}_{i j k l}$ is the Riemann tensor for the metric
$H_{i j}$ and $d\sigma_c = {d\sigma\over F}$. The first term in
equation \forty ~is connected with the time delay effect in a static 
gravitational field. The second non-diagonal term is connected 
with the curvature of the 3-dimensional unphysical space. We 
need to calculate the matrix potential to determine the time
dependent propagation of the perturbation. The normal vectors along
perpendicular and parallel to strings are chosen as,
$$
\eqalign{n^i_{\perp} & = {2\over{\sqrt{k_{S U}}}} (0, 1, 0)\cr
n^i_{\parallel} &= {2\over{\sqrt{k_{S U}}}} \left(- b, 0, 
\left({r^2\over l^2} - 1\right)^{-1} 
{d r\over{d\sigma_c}}\right)\cr}\eqn\fortytwo$$

Using equations \forty, \fortyo ~and \fortyone, we obtain,
$$
\eqalign{V &= P - {b^2\over\Delta}\left[P + \left({r\over l^2} - {J^2\over
{4 r^3}}\right)^2\right] \cr 
V_{\perp\perp} &= - P\left({r^2\over l^2} -1\right) + {4 b^2\over
{k_{S U}}}\left[P + \left({r\over l^2} - {J^2\over{4 r^3}}\right)^2 -
{1\over\Delta} \left({r^2\over l^2} - 1\right)^2\right] \cr
V_{\parallel\parallel} &= - P\left({r^2\over l^2} - 
1\right)\cr}\eqn\fortythree$$
where,
$$
P = \left({M\over l^2} + {3 J^2 M\over{4 r^4}} - {3 J^2\over{2 r^2 l^2}}
- {J^4\over{8 r^6}}\right)\eqn\fortyfour$$

The components of the matrix potential are given by,
$$
\eqalign{U_{\perp\perp} & = - {16 b^2\over{k_{S U}^2}}\cr
U_{\parallel\parallel} & = -{b^2\over{\Delta}}\left[P +
\left({r\over l^2} - {J^2\over{4 r^3}}\right)^2\right]\cr
U_{\perp \parallel} &= 0\cr}\eqn\fortyfive$$
where, $\Delta = {k_{S U}\over 4} \left(-M + {r^2\over l^2} + {J^2\over
{4 r^2}}\right)$. 
Now the equations of motion for the time-dependent perturbations 
in perpendicular and parallel directions are respectively given by,
$$
({\partial}_{\sigma_c}^2 - {\partial}_{\tau}^2)\delta x_{\perp} + 
{16 b^2\over{k_{S U}^2}} \delta x_{\perp} = 0\eqn\fortysix$$
and,
$$
({\partial}_{\sigma_c}^2 - {\partial}_{\tau}^2)\delta x_{\parallel}+
{4 b^2\over{k_{S U}\left({r^2\over l^2} - M + {J^2\over{4 r^2}}\right)}} 
\left[P + \left({r\over l^2} - {J^2\over{ 4 r^3}}\right)^2\right]\delta
x_{\parallel} = 0\eqn\fortyseven$$


For the straight string configuration, $b = 0$. So the perturbation
equations become identical and they reduce to plane wave equations,

$$
(\partial_{\sigma_c}^2 - \partial_{\tau}^2) \delta x_{\perp, \parallel}
= 0\eqn\fortyeit$$
By a Fourier expansion of $\delta x_{\perp,\parallel}$, we get,
$$
\delta x_{\perp, \parallel}(\tau, \sigma_c) = \int e^{- i \omega\tau}
D_{\omega}(\sigma_c) d\omega\eqn\fifftyone$$
The solutions of eqn. \fortyeit are given by,
$$
\delta x_{\perp , \parallel}(\tau, \sigma_c) = \int d\omega\left(
A_{\omega}^{
\perp , \parallel} e^{- i\omega(\tau - \sigma_c)} + B_{\omega}^{\perp ,
\parallel} e^{- i\omega (\tau + \sigma_c)}\right)\eqn\fiffty$$

Now for the circular string configuration (which is possible 
when the geometry does not have a horizon because of the 
condition $J > M l$), we have, ${4 b^2\over{k_{S U}}} = 1$. 
The perturbation equations in this limit are given by,
$$
\left(\partial_{\sigma}^2 - \partial_{\tau}^2\right) \delta x_{\perp}
+ {1\over b^2} \delta x_{\perp} = 0\eqn\fifftytwo$$
and,
$$
\left(\partial_{\sigma}^2 - \partial_{\tau}^2\right) \delta x_{\parallel}
- {4\over l^2} \delta x_{\parallel} = 0\eqn\fifftyth$$

With the Fourier expansion of $\delta x_{\perp , \parallel}$ (eqn. \fifftyone),
the above two rquations reduce to,

$$
{d^2\over{d\sigma_c^2}} D_{\omega} + \left[\omega^2 + {1\over b^2}
\right ] D_{\omega} = 0\eqn\fiffour$$
and,
$$
{d^2\over{d\sigma_c^2}} D_{\omega} + \left[\omega^2 - {4\over l^2}
\right ] D_{\omega} = 0\eqn\fiffive$$

 For the circular string configuration $\phi(\sigma_c) = {\sigma_c\over b}$,
where, $\sigma_c$ is periodic with period $2\pi b$. Since the potentials
for both the perturbations are constant, we can write down the solutions
as,
$$
\delta x_{\perp} = \int d\omega\left[A_{\perp} e^{-{i\over b} (\tau
{\sqrt{n^2 - 1}} \, \,\pm \,\,n\sigma_c)}\right]\eqn\fifftythre$$
and,
$$
\delta x_{\parallel} = \int d\omega\left[A_{\parallel} e^{-{i\over b}
(\tau{\sqrt{n^2 + {4 b^2\over l^2}}} \, \,\pm \,\,n\sigma_c)}
\right]\eqn\fiffour$$
where, $n$ is an integer. We note from here, that there is no 
unstable mode in the solution for $\delta x_{\parallel}$, whereas, 
unstable modes can arise in the solution of $\delta x_{\perp}$ 
when $n = 0$.  

\noindent

{\bf V. Conclusion}

In this paper, we have used the stationary string ansatz to 
study the exact solutions of string equations of motion and 
constraints in the background of exact, stringy four-dimensional 
dyonic black hole obtained by tensoring the two-dimensional
electrically charged black hole with ${SU(2)\over{Z(m)}}$ coset 
models. Since solving the equations for non-zero $J$ becomes 
quite complicated, we restrict ourselves to $J = 0$ and 
$M = 1$ limit of the above background to solve the geodesic 
equations in the three dimensional "unphysical" space with a 
metric $H_{i j}$. The straight string configuration is obtained
in terms of Weierstrass elliptic functions. We have also 
analyzed the possibility of having circular string configuration. 
Unlike the case of $2 + 1$ dimensional blackhole anti-de-Sitter 
background of Banados, 
Teitelboim and Zanelli, there can not be any circular string 
configuration in the above four-dimensional background if the 
geometry has to include a horizon. No multi
string solution  is found in this case. We have also studied the 
propagation of physical perturbations along the stationary string
in the given curved space-time by using a covariant formalism. 
The equations of motion for the time-dependent perturbations are
in the form of wave equations with a complicated matrix potential
term. For the straight string configuration (${\it i. e.}$ when the 
parameter 
$b$ goes to zero), the matrix potential becomes zero and the equations 
of motion for perturbations in 
both perpendicular and parallel directions reduce to simple plane
wave equations.  
The above method can also be applied to stringy
cosmological backgrounds to obtain various string configurations
and a systematic analysis of propagation of perturbation along the 
string configuration will be an interesting problem. To conclude,
this excercise is mainly an attempt to have a general understanding 
of the string dynamics in curved space-times. 

{\bf ACKNOWLEDGEMENTS:} 
I would like to thank Sayan Kar and K. Maharana for many useful
discussions.
A pool grant (no. 13-6764-A/94) from C. S. I. R. is acknowledged.
\refout 
\end